\documentclass{llncs}

\usepackage[utf8]{inputenc}
\usepackage{textcomp}
\usepackage[usenames,dvipsnames]{xcolor}
\usepackage{graphicx} 
\usepackage{booktabs}
\usepackage{microtype}
\usepackage{amsmath}
\usepackage{amssymb}
\usepackage{mathrsfs}
\usepackage{dsfont}
\usepackage{url}
\usepackage[font=small,labelfont=bf]{caption}
\usepackage[linesnumbered,ruled]{algorithm2e}

\definecolor{lgray}{gray}{0.6}

\newcommand{\Sol}[0]{{\cal X}}
\newcommand{\Neigh}[0]{{\cal N}}
\newcommand{\Real}{\mathop{\rm I\kern-.2emR}}

\begin{document}

\title{
A fitness landscape view on the tuning\\
of an asynchronous master-worker EA\\
for nuclear reactor design
}

\author{Mathieu Muniglia\inst{1} \and S\'ebastien Verel\inst{2} \and \\ 
Jean-Charles Le Pallec\inst{1} \and Jean-Michel Do\inst{1}}

\institute{
CEA (french Commissariat \`a l'Energie Atomique), France
\and
Universit\'e du Littoral C\^ote d'Opale,
LISIC, France
}

\maketitle

\begin{abstract}
In the context of the introduction of intermittent renewable energies, we propose to optimize the main variables of the control rods of a nuclear power plant to improve its capability to load-follow. The design problem is a black-box combinatorial optimization problem with expensive evaluation based on a multi-physics simulator. Therefore, we use a parallel asynchronous master-worker Evolutionary Algorithm scaling up to thousand computing units. One main issue is the tuning of the algorithm parameters. A fitness landscape analysis is conducted on this expensive real-world problem to show that it would be possible to tune the mutation parameters according to the low-cost estimation of the fitness landscape features.
\end{abstract}

\vspace{-0.8cm}
\section{Introduction}

In the actual context of energetic transition, the increase of the intermittent renewable energies contribution (as wind farms or solar energy) is a major
issue. On the one hand, the French government aims at increasing their part up to 30$\%$ \cite{Dumont:2012} by $2030$, against 6$\%$ today. On the other hand,
their intermittent production may lead to an important imbalance between production and consumption. Consequently, the other ways of production must adapt to
those variations, especially nuclear energy which is the most important in France. The power variations occur at different time scales (hour, day, or even week)
and in order to counterbalance their effects on the electric grid, the nuclear power plants (NPP) are able to adjust their production. NPPs which take
part in the response of the power variations operate in the so-called load-following mode. In this operating mode, the power plant is mainly controlled using
control rods (neutron absorber) that may introduce unacceptable spatial perturbations in the core, especially in case of huge power variations. The
purpose of this work is to optimize the manageability of the power plants to cope with a large introduction of intermittent renewable energies. Its final goal
is to tune the control parameters (called variables) in order to be able to make the load following at a shorter time scale and larger power amplitude scale,
meeting the safety constraints.

Such a real-world optimization problem is a challenge, considering the size of the search domain, the computation cost and the unknown properties of the
fitness function. Due to the design of the nuclear power plants, and in a goal to propose only simple modifications of the current management, $11$ integer
variables are used to describe the control rods such as speed, overlaps between rods, etc. (details are given in Sect. \ref{sec:problem}). Therefore, the
optimization is a large size combinatorial problem where no full enumeration is possible. Moreover, a multi-physic simulator is used which is able
to compute several criteria such as the evolution of the axial power offset, the rejected volume of effluent, etc. according to the variables of the
problem. So, the computation of the fitness function is computationally expensive, and one evaluation typically takes on average about $40$ minutes. This
optimization problem is considered for the first time, and no property on the search space is \textit{a priori} known. Hence, in this work, the Nuclear Reactor
Operation Optimization problem (NROO problem) is an original combinatorial black-box problem with expensive fitness function evaluation for which only few
candidate solutions and their corresponding fitness value can be computed.

The ability of Evolutionary Algorithm (EA) to find high quality solutions is likely to depend strongly on its parameters settings. In this work, we propose a
parallel master-worker EA for large scale computing environment to solve the NROO problem. Despite the expensive cost, an analysis of the mutation parameters is
then proposed. Such a study is not always possible for expensive optimization problems. Hence, we achieve a fitness landscape analysis of the NROO problem using
low-cost features to argue that it helps to select the relevant parameters of the mutation operator.

The main goals of the paper are then~: (i) perform for the first time an offline optimization of the control rods using an evolutionary algorithm (ii) analyze the
fitness landscape structure using a random walk (details are given in Sect. \ref{sec:FLA}) to tune the algorithm parameters, especially the ones of the
mutation, in order to (iii) propose an efficient mono-objective master-worker that will be used in the next step of the work, consisting in a multi-objective
optimization using a decomposition approach.

The rest of this paper is organized as follows. The next section introduces previous works on nuclear energy problems, and main definitions used in this work.
The NROO problem and the proposed algorithm are described respectively in Sect. \ref{sec:problem} and \ref{sec:algo}. The experimental analysis of the algorithm
and of the fitness landscape is conducted in Sect. \ref{sec:results}. At last, the paper concludes on the main results, and future works.

\section{Preliminaries}
\label{sec:related-works}

\subsection{Evolutionary optimization for nuclear energy problems}
The use of Evolutionary Algorithms (EA) in order to optimize some variables of a nuclear power plant as regards performance or safety is not new. Offline
optimizations can already be found, and studies such as \cite{Arnaud:2011} or \cite{Meneses:2010} deal with the In-Core Fuel Management
Optimization (ICFMO) and loading pattern optimization which is a well-known problem of Nuclear Engineering and aims for instance at maximizing the use of the
fuel (increase the cycle length for example) while keeping the core safe (minimize the power peak). Pereira and Lapa consider in \cite{Pereira:2003} an
optimization problem that consists in adjusting several reactor cell variables, such as dimensions, enrichment and materials, in order to minimize the average
peak-factor in a reactor core, considering some safety restrictions. This is extended in \cite{Sacco:2006} to stochastic optimization algorithms conceptually
similar to Simulated Annealing. Sacco \textit{et al.} even perform in \cite{Sacco:2008} an optimization of the surveillance tests policy on a part of the secondary
system of a Nuclear Power Plant, using a metaheuristic algorithm, which goal is to maximize the system average availability for a given period of time.

To our best knowledge, the only optimizations of the plant operation are made online, like in \cite{Na:2006}, where Na \textit{et al.} develop a fuzzy model predictive
control (MPC) method to design an automatic controller for thermal power control in pressurized water reactors. The objectives are to minimize both the
difference between the predicted reactor power and the desired one, and the variation of the control rod positions. A genetic algorithm is then used to optimize
the fuzzy MPC. Kim \textit{et al.} propose in \cite{Kim:2014} another MPC by applying a genetic algorithm, to optimize this time the discrete control rod speeds. This
paper proposes a new approach to do so, by optimizing offline the main characteristics of the control mechanisms, using an EA.
\vspace{0.2cm}

\subsection{Parallel evolutionary algorithms}
With the increasing number of computing units (cores, etc.), parallel EA become more and more popular to solve complex optimization problems. Usually, two main
classes of types of parallel EA \cite{Alba:2002} can be distinguished~: the coarse-grained model (island model) in which several EA share solutions within the
migration process, and the fine-grained model (cellular model) where the population is spread into a grid and evolutionary operators are locally executed.
Besides, a Master-Worker (M/W) architecture with the fitness evaluation on workers have been extensively used and studied \cite{Dubreuil:2006}. It is simple to
implement, and does not require sophisticated parallel techniques. Two communication modes are usually considered. In the synchronous mode, the parallel
algorithm is organized by round. The master sends candidate solutions on each worker for evaluation, and waits until receiving a response from all workers
before the next round. In the asynchronous mode, the master does not need to wait, and communicates with each worker individually on-the-fly. The asynchronous
mode could improve the parallel efficiency when the evaluation time of the fitness function vary substantially \cite{Wessing:2016}. We also propose an
asynchronous parallel EA in this work.

\subsection{Landscape aware parameter tuning}
\label{sec:fl}
The performance of EA strongly depends on the value of their parameters (mutation rate, population size, etc.). Parameters setting is then one of
the major issues in practice for EA, and two methodologies are commonly used \cite{Eiben:2007}. In the online setting, called \textit{control}, the parameters
are selected all along the optimization process. In the offline setting, called \textit{tuning}, the parameter values are set before the execution of the
algorithm. In offline setting, most of the methods, such as the irace framework \cite{Lopez-Ibanez:2011}, are based on a smart trial and error technic of
parameter values on a set of problem instances. Those methods may require a large number of tests/executions on representative problem instances which can be
difficult to afford in a black-box scenario with expensive costs on large scale computing environment. Alternatively, following Rice's
framework~\cite{Rice:1976}, one can use a fitness landscape aware methodology to first extract features from the given problem instance, then select the
relevant parameters according to those fitness landscape features.

Fitness landscapes are a powerful metaphor to describe the structure of the search space for a local search algorithm, and peaks, valley or plateaus for instance
are used to depict the shape of the search space in this picture. Formally, a fitness landscape \cite{Stadler:2002} is defined by a triplet $(\Sol, \Neigh, f)$
where $\Sol$ is the set of candidate solutions, $\Neigh : \Sol \rightarrow 2^{\Sol}$ is the neighborhood relation between solutions, and $f : \Sol \rightarrow
\Real$ is the fitness function (here assumed to be minimized) which associates to each candidate solution the scalar value to minimize. The neighborhood
relation can be defined by a distance between solutions or by a local search operator.

Two main geometries are commonly used in fitness landscape. A multimodal fitness landscape is a search space with a lot of local optima (solution with no
improving solution in the neighborhood). This geometry is also associated with the \textit{ruggedness} which is the local regularity of the landscape. The more
rugged the more multimodal the landscape is. The ruggedness can be measured by the autocorrelation of fitness \cite{Weinberger:1991} during a random walk over the
landscape. A random walk is a sequence $(x_1, \ldots, x_{\ell} )$ of solutions such that for all $t \in \{2, \ell \}$, $x_{t}$ is a neighboring solution
selected uniformly at random from $\Neigh(x_{t-1})$, or according to the local search operator. The autocorrelation function $\hat{\rho}$ is defined by the
correlation of fitness between solutions of the walk~:
$\hat{\rho}(k) = \frac{\sum_{t=1}^{\ell - k} (f(x_t) - \bar{f}) \cdot (f(x_{t+k}) - \bar{f})}{\sum_{t=1}^{\ell} (f(x_t) - \bar{f})^2}$
with $\bar{f}$ the average value of $f(x_t)$. The main feature of ruggedness is then the autocorrelation length \cite{Hordijk:1996} which is the length $\tau$ such
that there is no more significative fitness correlation at level $\epsilon$ between solutions of the walk~:
$\tau = \operatorname{min} \{ k ~:~ | \hat{\rho}(k) | < \epsilon \}$. Usually, a smooth fitness landscape with long autocorrelation length is supposed to be
easier to solve.

A neutral fitness landscape is another main geometry where the search space is dominated by large flat plateaus with many equivalent solutions. The dynamics of
EA on such landscape is characterized by punctuated equilibrium dynamics where long neutral moves on plateaus are interrupted by rapid improving moves toward better
solutions. One of the main features of this landscape is the neutral rate which is the proportion of neighboring solutions with the same fitness value
\cite{Vanneschi:2007}~: $\text{E}_{\Sol}[ \sharp \{ y ~:~ f(y) = f(x) \text{ and } y \in \Neigh(x) \} / \sharp \Neigh(x) ]$. To avoid the computation of large
neighborhood, the neutral rate can be estimated with a random walk \cite{Liefooghe:2017} by~: $nr = \sharp \{ (x_t, x_{t+1}) ~:~ f(x_t)=f(x_{t+1}), t \in
\{1, \ell-1 \} \} / (\ell - 1)$.

According to the local search operator, which could be the mutation operator for EA, the features of fitness landscape can characterize the shape of the
landscape. First fundamental works have demonstrated the relevance of fitness landscape analysis for the parameters tuning \cite{Daolio:2016}. However, to our
best knowledge, no work has used such methodology for a real-world problem with expensive fitness function.

\section{Problem definition}
\label{sec:problem}

\vspace{-0.1cm}
The optimization process is based on the current load-following transient \cite{Lokhov:2011} and this analysis focuses on a single Pressurized Water Reactor
(PWR) type (1300 MW) of the French nuclear fleet. When an electrical power variation occurs (demand of the grid) a chain of feedback is setting up in the whole
reactor, leading to a new steady state. It is usual to take advantage of this self-regulation in the case of small variations, but the regulated variables such
as the temperature or the pressure in the primary or secondary circuits may reach unacceptable values in case of load-following, possibly leading to damages of
the whole system. The control rods are then used in order to cope with this variation, and maintain the primary coolant temperature close to the target.
However, those control rods have to be handled carefully as they could cause axial or radial heterogeneity in the core, inducing high power peaks or Xenon 
oscillations.

\vspace{-0.2cm}
\subsection{Description of the system}

\begin{figure}[t!]
\centering
\centering
  \includegraphics[width=0.9\textwidth]{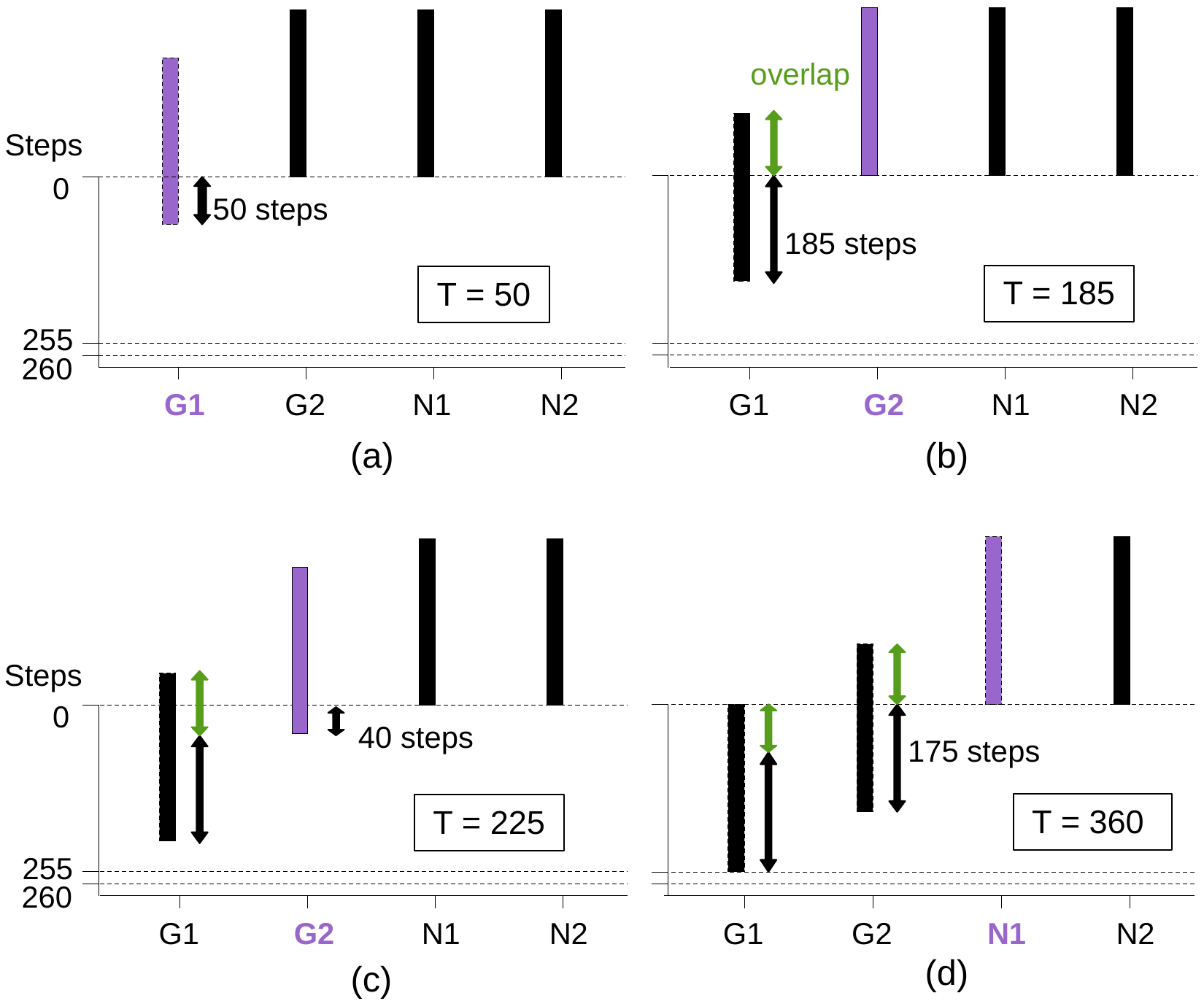}
  \caption{Insertion sequence of the Power Shimming rods (PS). The totalizer value ($T$) is given on each frame, and the last
moving group is in purple.}
  \label{fig:PSR}
\end{figure}

\begin{figure}[t!]
\centering
  \includegraphics[width=0.9\textwidth]{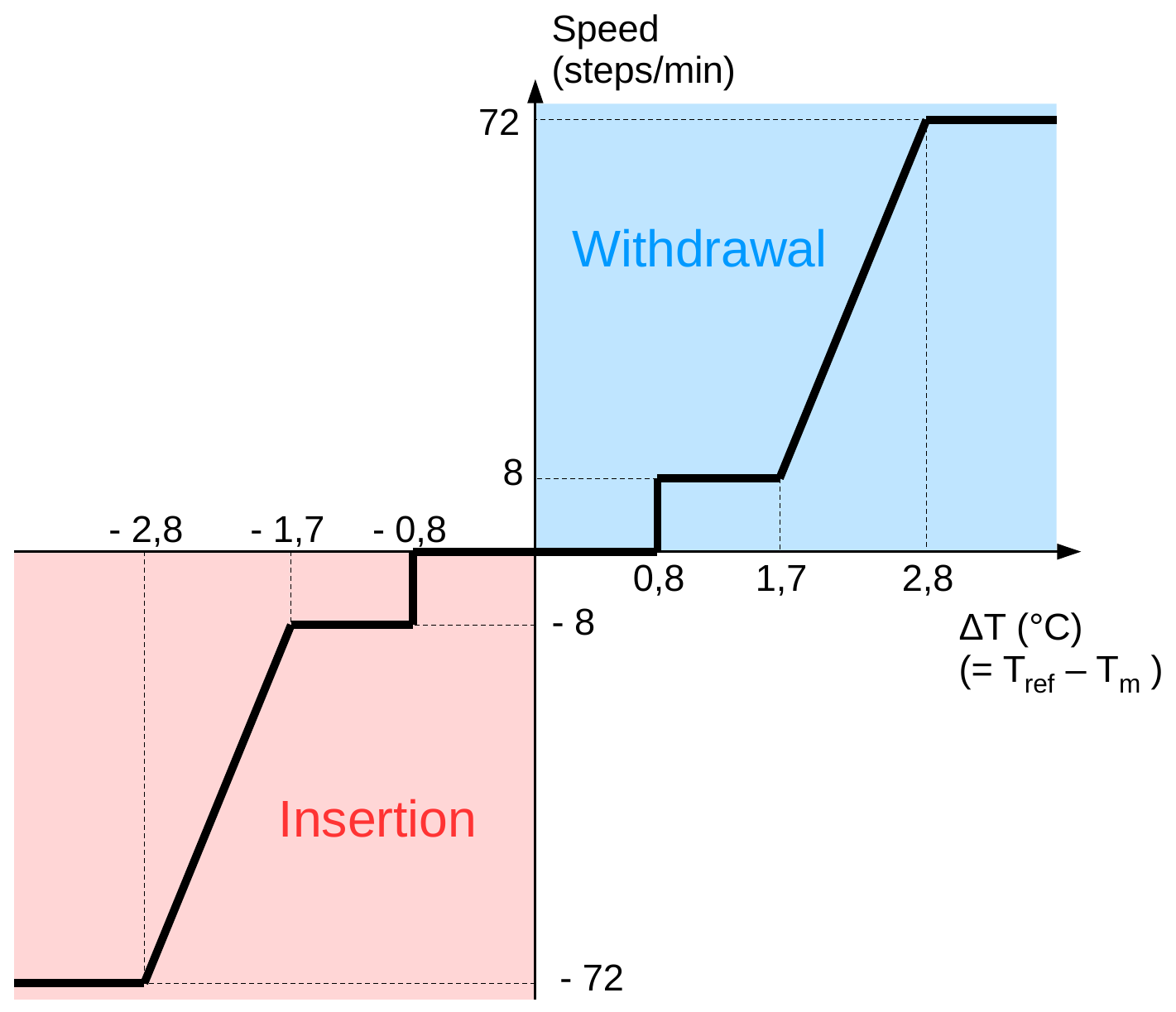}
  \caption{Speed program of the Temperature Regulation rods (TR). The dead band corresponds to the null speed and the maximal and minimal
speeds ($\pm 72$ steps/min) are for an absolute temperature difference larger than $2.8$\textcelsius.}
  \label{fig:TRR}
\end{figure}

The reactor core is a grid of square assemblies (21cm length) in a cylindrical vessel. 
There are $193$ assemblies, split into two kinds : $120$ assemblies made of Uranium oxide ($UOX$) and $73$ ones made of Uranium plus Gadolinium oxides ($UGd$). 
Each control rod is made of pins of a neutron absorber that are inserted together from the top of the core in some assemblies. The positions of the assemblies 
where they are inserted and the materials of which they are made correspond to the French “G” mode \cite{Lokhov:2011}. The rods are organized in two families: (i) the power shimming rods (PS) and (ii) the regulation rods (TR).
The first ones are used to shim the power effects during the power transient, and are split in four groups (4 rods G1, 8 rods G2, 8 rods N1 and 8 rods N2). All
the rods of a same group move together, and the groups are inserted successively in this order~: G1, G2, N1, N2, as it is shown in Fig.\ref{fig:PSR}. An
overlap is also defined between all the groups, so that they follow an insertion program as illustrated from frames $(a)$ to $(d)$. The position of those rods 
is linked to the electrical power by a calibration function.
The second family enable a control of the average coolant temperature of the core (the targeted temperature, called reference temperature, is a linear
function of the thermal power) and is made of 9 rods gathered in a single group. This group moves independently and automatically, following a speed program 
depending on the difference between the reference temperature ($T_{ref}$) and the mean temperature ($T_m$) as shown in Fig.\ref{fig:TRR}. One can see a dead 
band of $\pm 0.8$\textcelsius\ in which the rods do not move, avoiding continuous displacement and corresponding to the self-regulation of the core. Finally, as
they are very efficient and for safety reasons, they are shut into a maneuvering band of about $50$ centimeters in the upper part of the core. For more details, please
refer to \cite{Grard:2014}. 

The variables to be tuned for the optimization are then the $4$ nominal speeds and the $3$ overlaps for the PS rods, the maximal and minimal speeds, the dead
band width and the maneuvering band height for the TR rods. $11$ variables are then considered, and they are coded as integer values corresponding to a discrete
number of steps or of temperature (the dead band is discretized by steps of $0.1$\textcelsius). Table~\ref{tab:sum_par} summarizes the variables, their initial
values (current management) and ranges. 
The values take into account some technological and logical constraints. For example, the overlaps cannot be greater than the total height of the 
rods, the velocity ranges are bounded by the mechanisms, etc. A number of other variables could have been studied, like swaps between groups, or splitting 
groups, but the study is confined to the variables listed for two reasons: simplify the problem for a first optimization, and be able to propose a solution 
without major technological breakthroughs and similar to the current one. Nevertheless, the search domain is huge (at least $3 \times 10^{20}$ possible configurations). 

\begin{figure}
  \centering
  \includegraphics[width=0.5\textwidth]{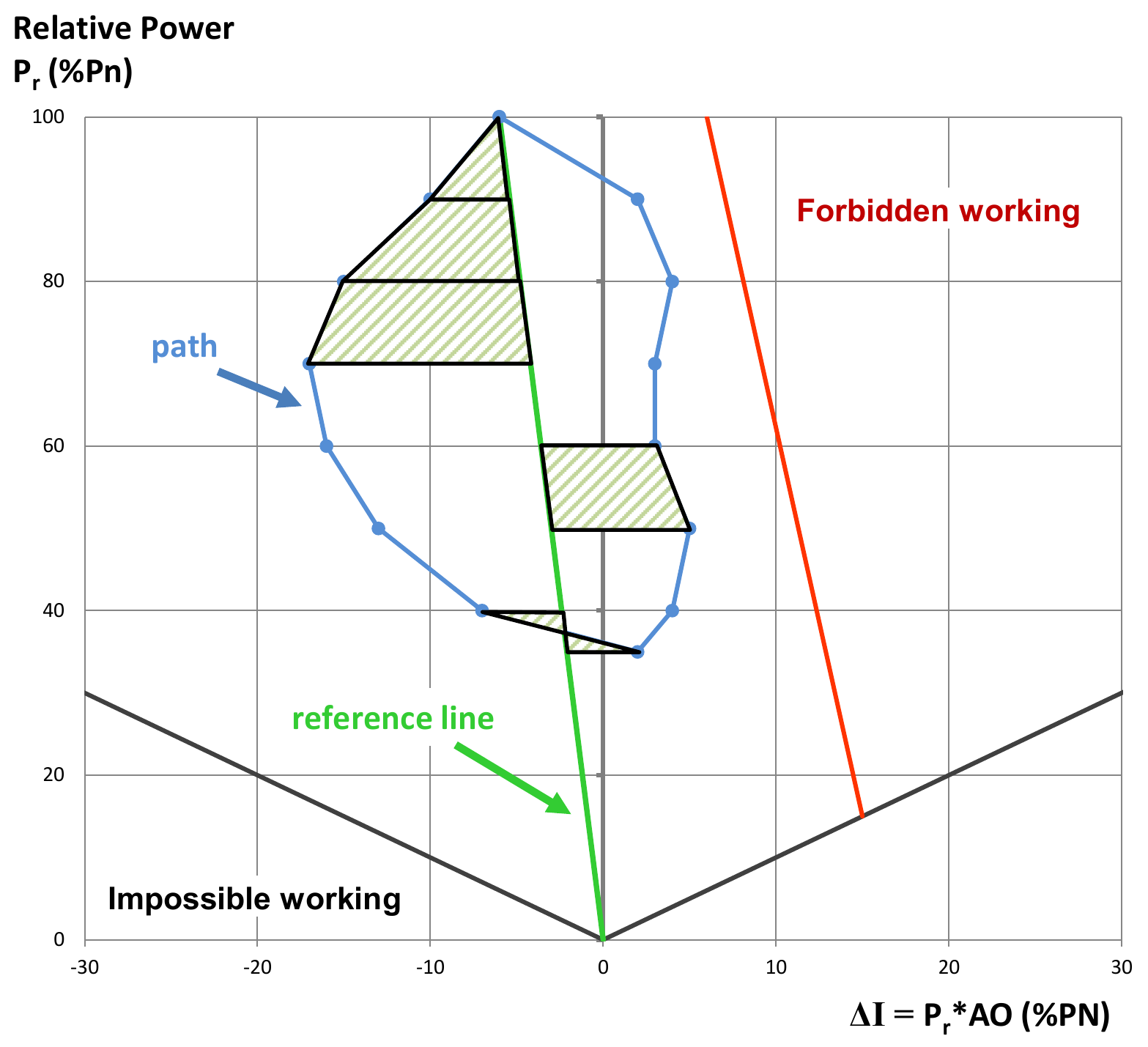}
  \caption{Control diagram and criteria calculation principle.} 
  \label{fig:control_diagram}
\end{figure}

\begin{table}
  \centering
  \caption{Integer variables of the design: lower bound (l.), upper bound (u.), and value of the current reference (r.).
The dead band (db) variable is expressed in tenth of degree, and all the other variables are expressed in steps.}
  \label{tab:sum_par}
  \begin{small}
\begin{tabular}{|p{0.3cm}|p{0.5cm}|p{0.5cm}|p{0.5cm}|p{0.5cm}|p{0.5cm}|p{0.5cm}|p{0.5cm}|p{0.5cm}|p{0.5cm}|p{0.5cm}|p{0.5cm}|}
\hline
& \multicolumn{3}{|c|}{PSR Overlaps} & \multicolumn{4}{|c|}{PSR Velocities} & \multicolumn{2}{|c|}{TRR V.} & \multicolumn{2}{|c|}{} \\
\hline
& $o_1$ & $o_2$ & $o_3$ & $v_1$ & $v_2$ & $v_3$ & $v_4$ & $V$ & $v$ & mb & db \\
\hline
\hline
l. & $0$ & $0$ & $0$ & 10 & 10 &  10 &  10 & 3 & 3 & 7 & 8 \\
\hline
u. & $255$ & $255$ & $255$ & 110 & 110 & 110 & 110 & 13 & 13 & 117 & 16 \\
\hline
\hline
r. & $185$ & $175$ & $160$ & $60$ & $60$ & $60$ & $60$ & 72 & 8  & 27 & 8 \\
\hline
\end{tabular}
  \end{small}
\end{table}

\subsection{Criterion of interest}

This seek of simplification is even more understandable when it is known that the black-box evaluation function is very costly. Each unitarian calculation 
corresponds to a given management configuration running on a complete typical load-following transient, corresponding to about $11$ hours. The value of
interest is then determined thanks to a model of the whole reactor described in \cite{Muniglia:2016}, and developed within
the APOLLO3\textregistered~\cite{Schneider:2016} calculation code. The optimization aims at minimizing this value of interest, which represents a global
operating criterion, based on the control diagram. This control diagram is used by the operator to manage the power plant and represents the evolution of the
relative thermal power ($P_r$) as a function of the power axial imbalance given by: $\Delta I = P_r \times AO$
where $AO$ is the axial offset defined as $AO = \frac{P_T - P_B}{P_T + P_B}$ and standing for the unbalance between the lower and upper half parts of the core
as regards the power. $P_T$ (resp. $P_B$) is the power in the upper (resp. lower) part of the core.

An example of such a diagram is to be found on Fig.\ref{fig:control_diagram}, which draws the path of the state of the core during a power 
variation (blue line) and the bounds for this path. On the right side, the forbidden region (red line) is based on many studies and ensures the
safety of the core in case of accidental situations. The impossible working region just comes from the definition ($AO \in [-1,1]$). Finally, the green
line starting at the same point as the path corresponds to a constant axial offset, and is called reference line in the following.
The criterion derived from the control diagram to be minimized is defined by:
\begin{equation} \label{criterion}
\vspace{-0.5cm}
f(x)= \frac{1}{4} \sum_i | P_{r,i+1}^2 - P_{r,i}^2 | \cdot \Big( D (\Delta I_{i+1}) + D (\Delta I_{i}) \Big)
\end{equation}
where $D (\Delta I_{i}) = | \Delta I_{i} - \Delta I^{ref}_{i} |$. The pair ($P_{r,i}$, $\Delta I_{i}$) represents the state of the core at the time step $i$, and $\Delta I^{ref}_{i}$ the power axial imbalance given by the reference line at the power $P_{r,i}$. The criterion corresponds to the sum of all the areas as illustrated on Fig.~\ref{fig:control_diagram}, weighted by the relative power to take into account the fact that an important axial offset at high power is worse than at low power. Minimizing this criterion enables to reduce the area of the path and avoids being close to the forbidden region while staying close to the reference line.

\vspace{-0.3cm}
\section{Asynchronous parallel EA}
\label{sec:algo}

\vspace{-0.2cm}
The design of the EA is guided both by the expensive cost of fitness evaluation of the problem computed by a numerical simulation, and by the computing 
environment available to solve this problem. 

 \subsection{Algorithm definition}

On the one hand, the fitness evaluation duration is about $40$ minutes on average with a large variance. On the other hand, a large number of computing units 
($w=3072$) are available to run the optimization algorithm, but they are only free for few hours (around 15 hours per experiment). Hence, we propose a 
master-worker (M/W) framework for the EA. On average one fitness evaluation is completed every $0.78$ second, meaning that the master node is not to be 
overflowed by the request of the workers, and with respect to the fitness evaluation time, an idle working time of few seconds will not reduce the 
performance. In addition, some simulations crash before the end of the calculation, increasing even more the discrepancies in calculation times. All 
considered, the model of the M/W has been made asynchronous: the workers are updated on the fly without a synchronization barrier, and each worker only 
computes the fitness value using the multi-physic simulator.

A lot of efficient EA can be considered in an asynchronous M/W framework with fitness evaluation on workers. The number of evaluations per worker is small, on 
average $23$ fitness function evaluations is possible on each worker within $15$ hours of computation. As a consequence, the EA should converge quickly. We 
propose then an asynchronous $(1+\lambda)$-EA where $\lambda$ is the number of computation units minus one. The Algo.~\ref{algo:base} show the details of the algorithm.

\begin{algorithm}
          \For{$i$ in Workers} {
               $x^i \gets$ \textbf{Initialization} using quasi-random numbers \\
               \textbf{Send (non-blocking)} Msg($x^i$) to worker $i$ \\
          }
          $f^{\star} \gets $ maximal value \\
          \While{pending message \textbf{and} time is not over}{ 
               \textbf{Receive} Msg from worker $i$ \\
               $f^i \gets $ Msg[0] \\
               \If{$f^i \leqslant f^*$}{
                   $x^{\star} \gets x^i$ ; $f^{\star} \gets f^i$\\
           }
               $x^i \gets$ \textbf{Mutate}$(x^{\star})$ \\
               \textbf{Send (non-blocking)} Msg($x^i$) to worker $i$ \\
          } 
    \KwRet{$x^{\star}$}
     \caption{Asynchronous M/W $(1+\lambda)$-EA on master}
     \label{algo:base}
\end{algorithm}

First, the algorithm on master node produces $\lambda=w-1$ quasi-random solutions (integer vectors of dimension $n=11$) using a Design of Experiments (DoE)
based on Sobol of quasi-random numbers. This initialization is used to improve the spreading of the initial solutions in the search space. Every initial
solution is then sent asynchronously to a worker who
receives the solution from the master, computes the fitness value by
running the multi-physic simulator, and send back the result to the master node. In the meantime, the main loop of the Algo. \ref{algo:base} is executed on the
master node~: wait for a message from a worker $i$, and when the fitness value is received, the best so far solution is updated if necessary. Notice that the
best solution is replaced by the new solution evaluated by the worker even when the fitness values are equals. In that way, the algorithm is able to drift on
plateaus of the search space. A new candidate solution is then computed by the mutation (detailed in the next section) of the best-known solution and sent in
non-blocking mode to the same worker $i$. The master is then able to manage the requests of the other worker nodes by the asynchronous communication
mode. The algorithm stops after an arbitrary time limit is reached.

 \subsection{Mutation operator}

The mutation operator is based on the classical mutation for vectors of numbers. The mutation rate $p$ defines the parameter of the Bernouilli distribution to
modify each number of the vector. Therefore, the number of modified variables follows a binomial distribution of parameters $n$ and $p$, and the expectation of
the number of modified variables is $np$. When an integer variable is modified according to the mutation rate, a random integer number is drawn using a uniform
distribution centered on the current value. Let $x_j$ be the current value of the variable $j$, and $\delta_j$ the gap defined by $\lfloor r . (ub_j-lb_j) 
\rfloor$ where $lb_j$ and $ub_j$ are respectively the lower bound and the upper bound of the variable $j$ defined in the Tab. \ref{tab:sum_par}, and $r \in [0,
1]$ is a mutation parameter. The new value of variable $j$ after mutation is selected uniformly in the interval $[ x_j - \delta_j, x_j + \delta_j ] \cap [lb_j,
ub_j] \setminus \{ x_j \}$. The parameter $r$ tunes the range width for the new value of variable after the mutation, and is expressed relatively to the total
range width of the variables ($r \leq 0.5$).

In addition, to avoid multiple costly evaluations of the same candidate solution, a hash-map is used on the master node to save all evaluated solutions. The
mutation is applied on the solution until a new candidate solution which is not in the hash-map is produced by the mutation random process.

\section{Experimental analysis}
\label{sec:results}

First, the performance of the algorithm with a baseline parameters setting is studied with $3072$ computing units during $24$ hours (approx. $73,728$ hours of
CPU time). Then, the mutation parameters are analyzed with the algorithm launched on $3072$ computing units during $5$ hours (approx. $15,360$ hours of
CPU time per run). At last, a fitness landscape analysis is conducted. 

\vspace{-0.3cm}
\subsection{Baseline parameters setting}

Following the value of the mutation rate parameter of $1/n$ commonly used in EA, the mutation rate has been set roughly to the inverse of the number of variables ($p_0 =
0.1$), so that the mutation operator modify on average one variable. The width of the random variation range has been arbitrarily set to about $r_0 = 0.05$
($5\%$ of the total variation range of the variable). Those parameters have been chosen for the first optimization process and are called in the following the
\textit{baseline} settings. The use of an asynchronous
algorithm to avoid idle time is justified by the discrepancies of the computation costs from a candidate solution to another one. The mean computation time is
$2426$ seconds, and the faster computation is done in $1629$ seconds whereas the longer is performed in $6169$ seconds. Fig.\ref{fig:dynamic} shows the dynamic
of the run. The normalized best fitness
is drawn as a function of the number of evaluations received by the master node. A point is plotted when the best solution so far is updated (included for equal
fitness values). The fitness values are normalized by the fitness value of the current management (see Tab. \ref{tab:sum_par}). The solutions for which the
number of evaluations is lower than $3072$ are from the initial quasi-random population. Even if the best solution obtained with \textit{baseline} settings enable to reduce the fitness of
about $40\%$ compared to the current management, it can be seen that the number of strictly improving solutions is low (about $10$ improving steps). The
dynamic is a punctuated equilibrium dynamic with a lot of neutral moves on plateaus, and few improving solutions. For instance, the process is stuck on a
plateau at the end of the run~: almost $50,000$ fitness evaluations are necessary to find a strictly better solution. Subsequently, one can say
that the neutrality is really important in the NROO problem. This first experiment shows the relevance of the algorithm to found better solutions than the
current management, but it suggests that the setting of mutation parameters could also be improved.

\vspace{-0.2cm}
\subsection{Impact of the mutation parameters}

\vspace{-0.2cm}
This section deeply analyzes the influence of the mutation parameters on the performance of the M/W algorithm. Four values of mutation rates $p$ and
mutation ranges $r$ are investigated: $p \in \{0.1,0.2,0.3,0.4\}$ and $r \in \{0.05,0.1,0.2,0.5\}$. All the combinations are considered, given $16$ possible
mutation settings of the mutation operator. To reduce the intrinsic random effect of the algorithm, each couple of mutation parameters values $(p,r)$ have been
launched five times with different initial populations generated by the Sobol sequence of quasi-random numbers. However, the 5 initial populations are the same
for each couple of parameters settings. The total computation cost is more than $1,2 \times 10^6$ hours of computation times, and we were not able to execute
more than five runs.

\begin{figure}[t!]
\centering
\includegraphics[width=0.9\textwidth]{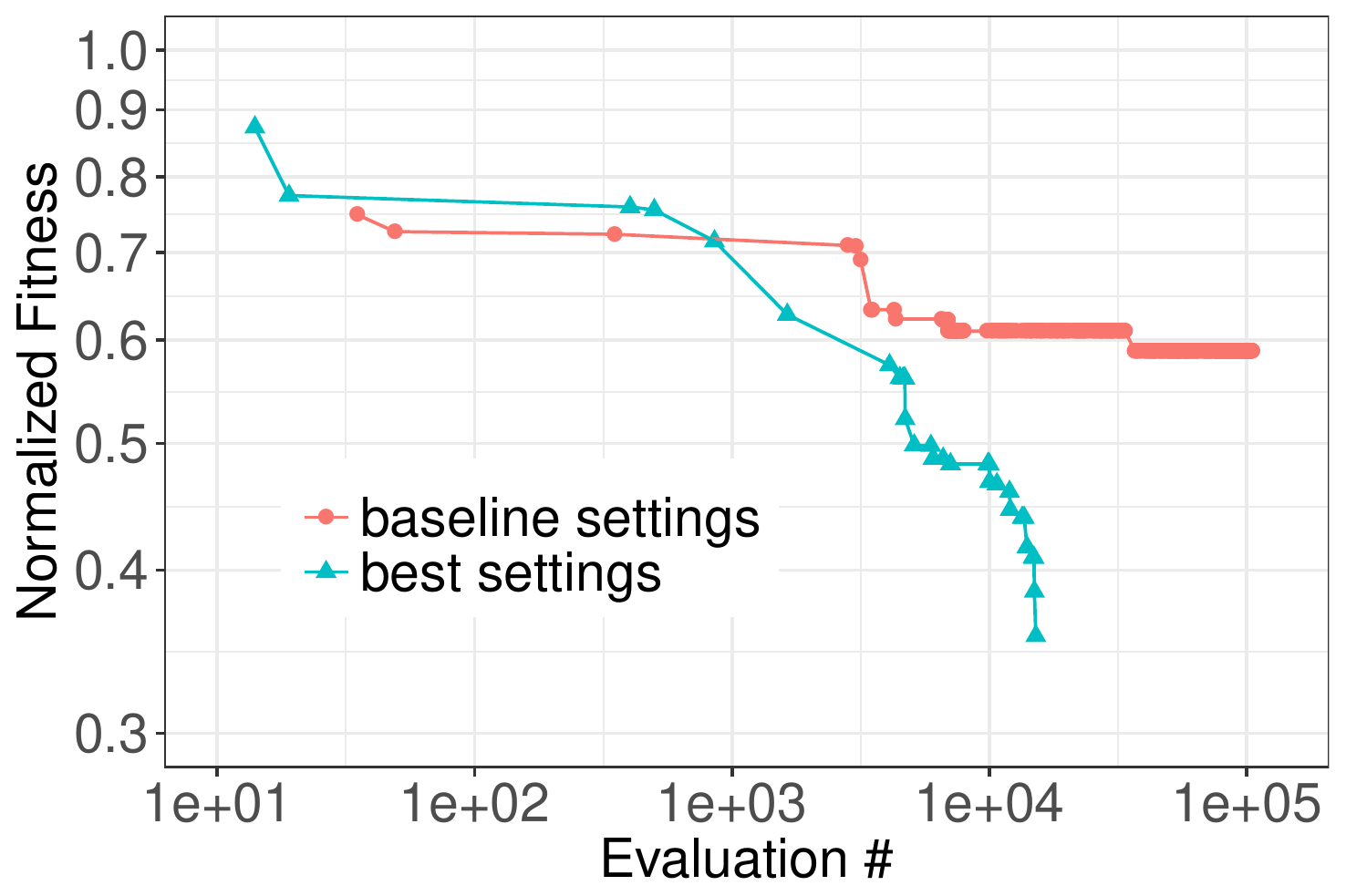}  \\
\caption{Dynamic of the asynchronous M/W algorithm for the \textit{baseline} and optimal mutation parameters settings} 
\label{fig:dynamic}
\end{figure}

\begin{figure}
\centering
  \includegraphics[width=0.95\textwidth]{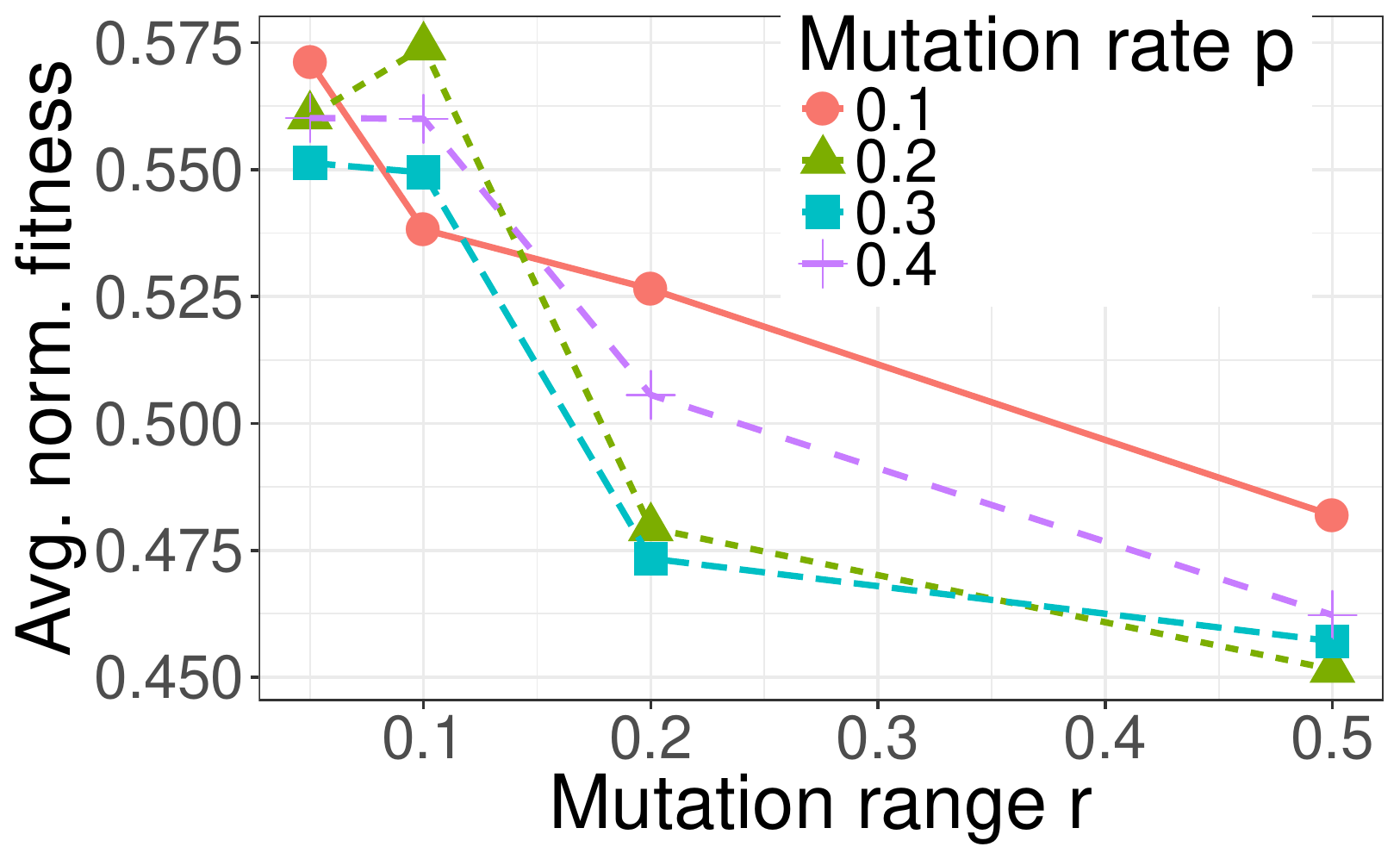}
\caption{Average normalized best fitness as a function of the mutation range width $r$ and mutation rate $p$.}
\label{fig:perf_optim}
\end{figure}

The Fig.~\ref{fig:perf_optim} shows the average normalized best fitness found for each parameter setting. The standard deviation of the best fitness found is
also computed to measure the robustness of the parameters settings (not shown here to save space). In addition, for each
initial population, the rank of each parameter setting is computed, and the average of the ranks gives another performance measure of the settings. However, 
statistical tests will not give exploitable results because of the very low number of runs, and are then not considered. The variation of average fitness is 
larger according to the mutation range width parameter $r$ than according to the mutation rate parameter $p$. The average fitness decreases with the parameter 
$r$ whereas there is no clear trend as a function of the mutation rate $p$. The best sets as regards this criterion are then the ones for which the mutation 
range $r$ is maximal. Inversely, the worse are the one for which the mutation range is minimal. Given the huge discrepancies of the average fitness as a 
function of the mutation range, the impact of the mutation rate cannot clearly be seen in this figure, and it is then difficult to choose the best mutation 
rate.

The performance according to the rank instead of best fitness value share the same result. Indeed, the Spearman correlation between the average fitness and the
mean rank appears to be really high ($\rho = 0.91$), meaning that the best parameters as regards the first one is likely to be good also as regards the
second. For example, the first five parameters settings with respect to the best average fitness are $(0.5, 0.2)$, $(0.5, 0.3)$, $(0.5, 0.4)$, $(0.2, 0.3)$,
$(0.2, 0.2)$, 
and they are respectively third, first, sixth, second and fourth with respect to the average rank. However, the correlation between the average and the standard
deviation of best fitness is low ($\rho = 0.48$) and thus, the five previous parameters settings are now in first, ninth, second, fifth and thirteenth position
with respect to the standard deviation. It was decided to prefer the mutation parameters leading to low fitness value and rank rather than to low standard
deviation. Future works will investigate ways to improve the robustness of the algorithm with respect to the initial population and thus to reduce the standard
deviation.

The selected parameters setting is then $r=0.5$, and $p=0.3$ which is the first (resp. second) with respect to the rank (resp. best average fitness) because
$(r,p)=(0.5, 0.2)$ is the first one as regards the best average fitness is only third as regards the rank, and also because the very best fitness so far is
obtained with $(0.5, 0.3)$. In the framework of the greedy $(1+\lambda)$-EA with large $\lambda$ value and low numbers of iterations, very large mutation
parameters with large exploration seem to be suggested.
The dynamic of the optimal parameters setting is shown on Fig.~\ref{fig:dynamic}. On the contrary of the common value of mutation parameters, the search is not
stuck on plateaus, and the number of improving steps is high. Besides, those parameters setting found an optimal solution which reduces almost $65 \%$ of the
reference fitness of current management, with only the quarter of the computation cost of the baseline settings.

\subsection{Fitness landscape analysis}
\label{sec:FLA}

In this section, we investigate the fitness landscape of the NROO problem. For each parameters setting of the mutation operator, a random walk of length $\ell
=1024$ starting from a random candidate solution is computed. The cost of the walk is about $5 \%$ of the computation cost of the EA, and the length is smaller
than the initial population size. Notice that by construction, all the solutions of the walk are strictly different. From the random walks, the autocorrelation
length and the neutral rate are both estimated (see Sect. \ref{sec:fl}). The significant level $\epsilon$ used to estimate the autocorrelation length is set to
$4/\sqrt{\ell}$. 

\begin{figure}[t!]
\centering
        \includegraphics[width=0.45\textwidth]{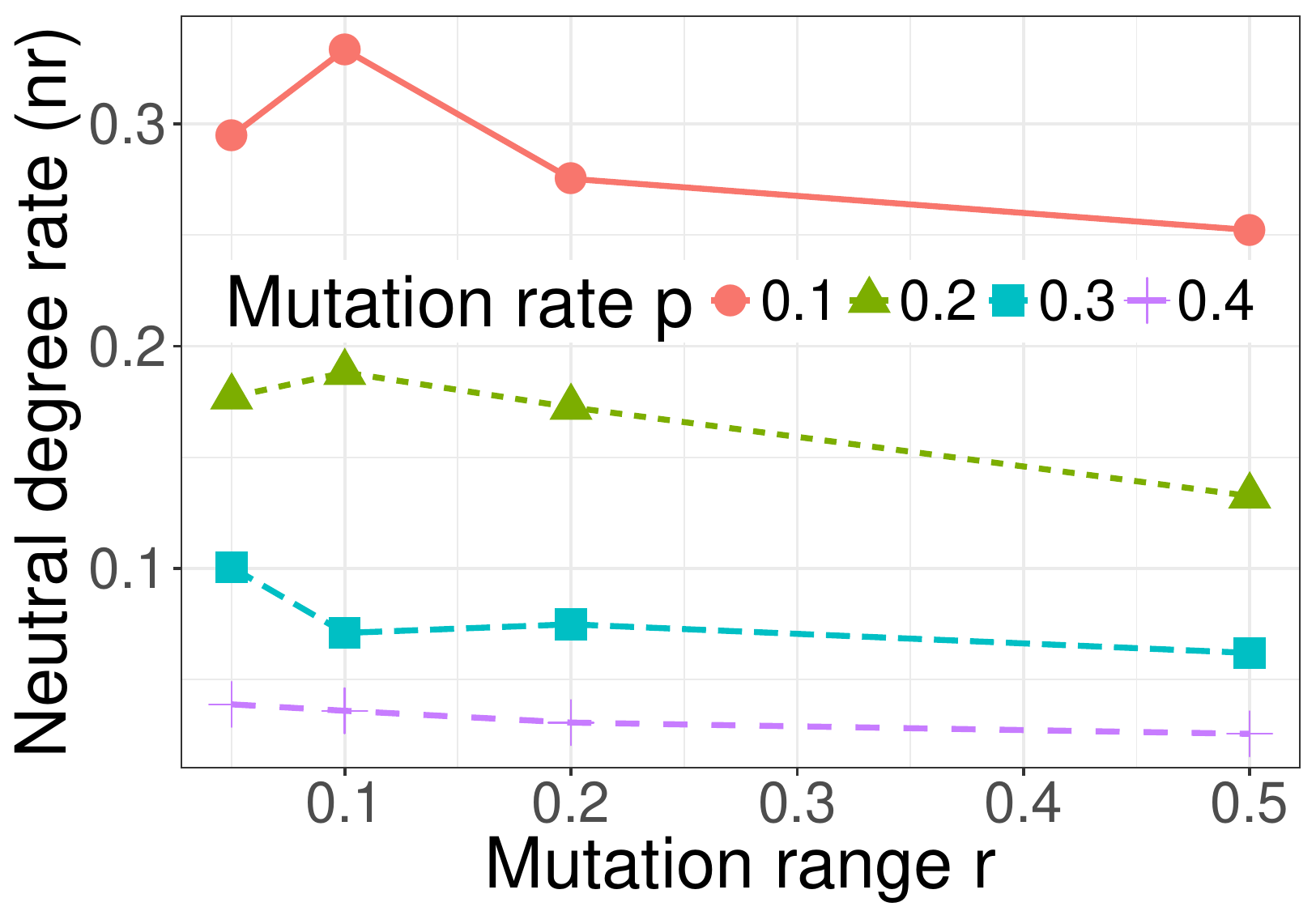}
        \includegraphics[width=0.45\textwidth]{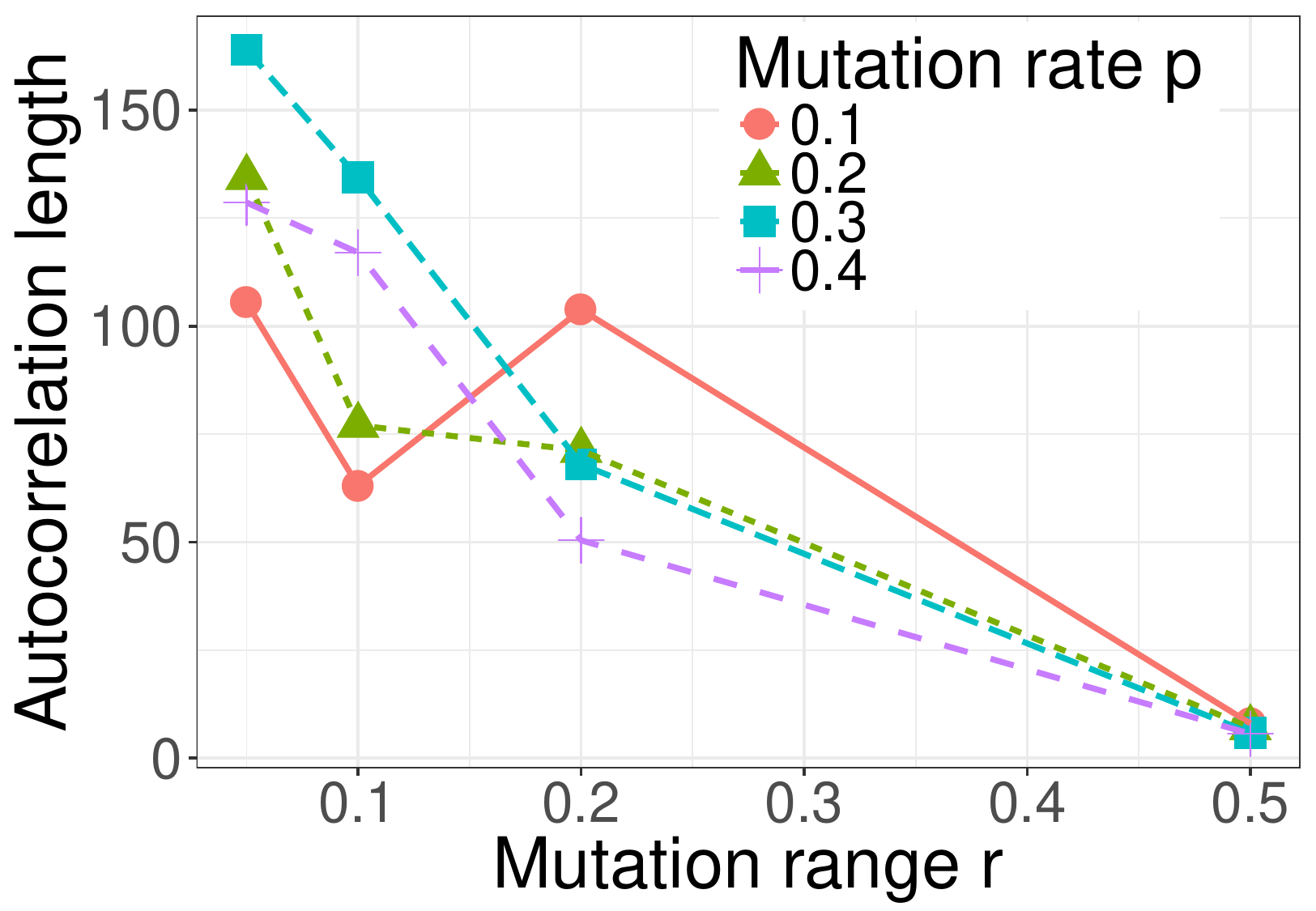}
\caption{Features of the fitness landscape as a function of mutation parameters $r$ and $p$. neutral rate (left) and autocorrelation length (right).}
\label{fig:features}
\end{figure}

The Fig. \ref{fig:features} shows the features of the fitness landscape according to the mutation parameters. The mutation range width $r$ does not impact 
the neutral rate. On the contrary, the neutral rate decreases with the mutation rate $p$~: from $25 \%$ for the \textit{baseline} setting with $p=0.1$ to $3\%$
for a high mutation rate value $p=0.4$. The neutrality of NROO fitness landscape is high, and is dominated by large plateaus for common value of the mutation
rate $p$. The neutral geometry explains the punctured equilibrium dynamics of the EA. As expected, a stronger mutation implies a more rugged fitness landscape.
However, the ruggedness of the landscape is more impacted by the mutation range width $r$ than by the mutation rate $p$. The autocorrelation length decreases
with the mutation range width $r$ from approximately $120$ for $r=0.05$ to $6$ for the largest value $r=0.5$ which picks a random new value. However, the
landscape can be considered as a smooth landscape. For instance, when the mutation range is $r=0.2$, more than 50 steps are required to reach a correlation of
fitness between solutions smaller than $\epsilon = 0.125$. This feature should explain the good performances of the EA.

\begin{figure}[t!]
\centering
        \includegraphics[width=0.45\textwidth]{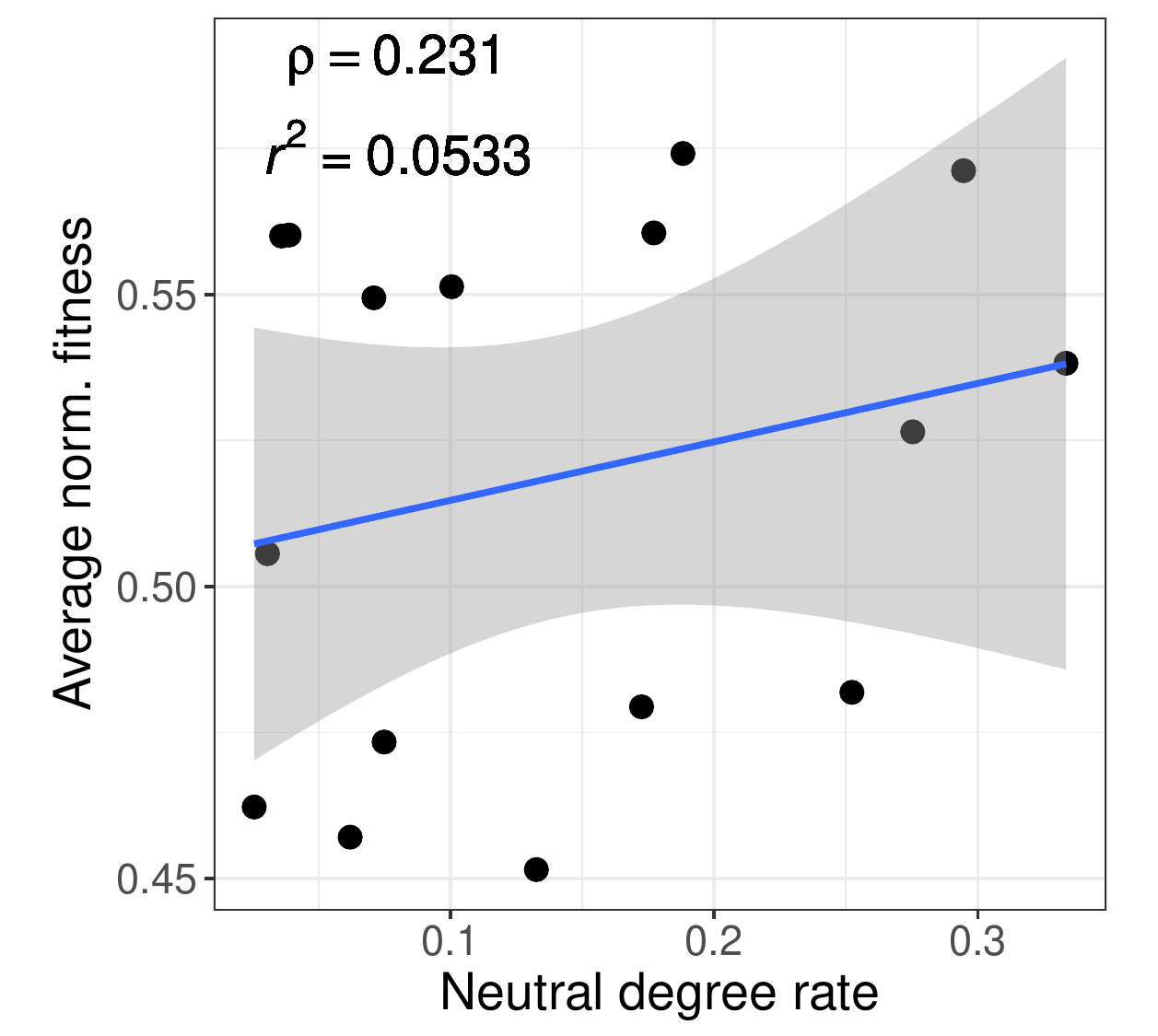}
        \includegraphics[width=0.45\textwidth]{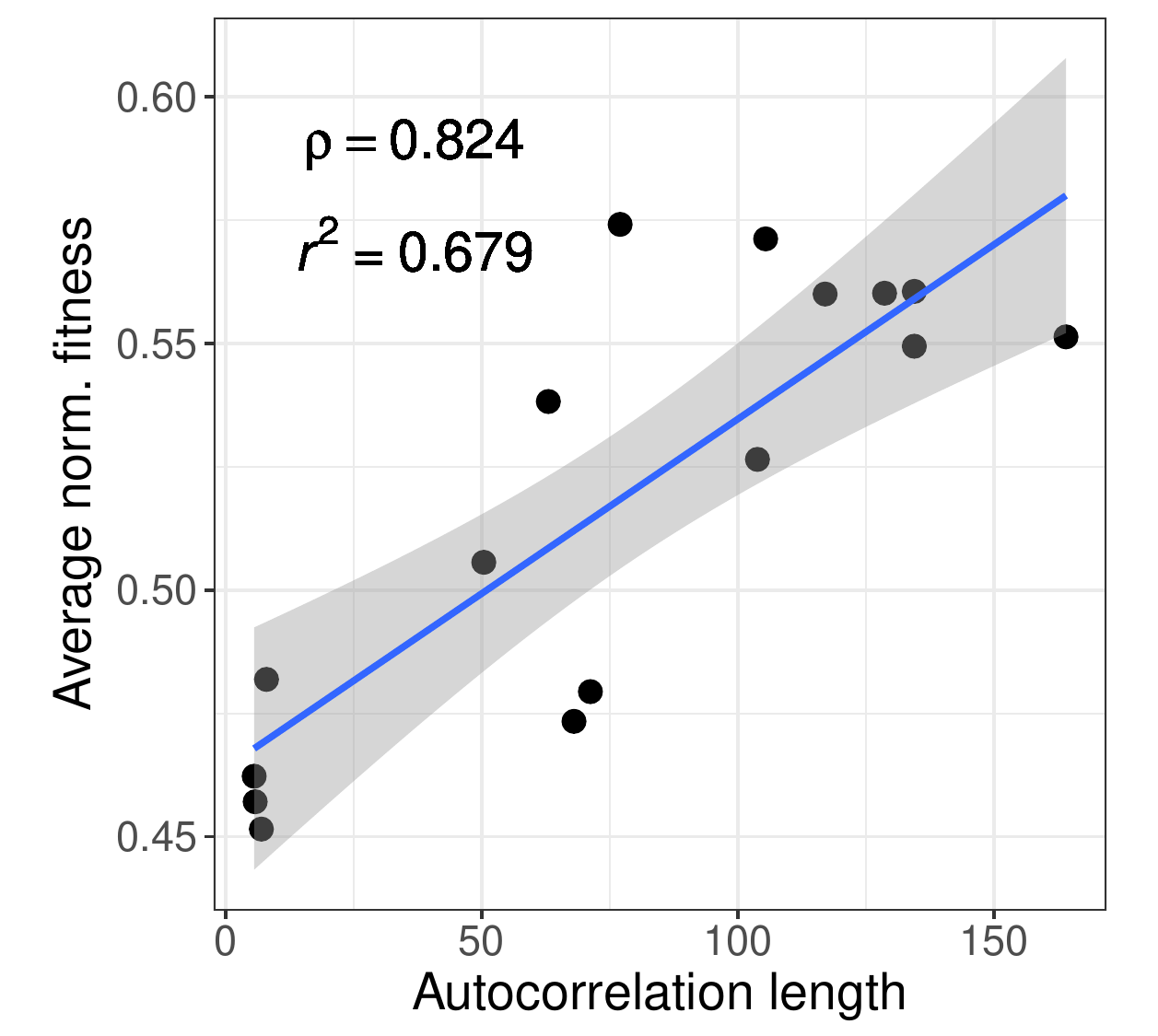}
\caption{Scatter plots and regression linear models between the average best normalized fitness and the features of fitness landscapes. Neutral rate (left) and autocorrelation length (right).}
\label{fig:corr}
\end{figure}

The Fig. \ref{fig:corr} shows the correlation between the performance of the EA in terms of average normalized best fitness found and the feature values of the 
fitness landscape. Each point corresponds to a mutation parameters setting and the regression line of the linear model is also drawn. Surprisingly, although the
neutral rate could be high, it is not linearly correlated to the performance of the EA. Only $r^2=5.3 \%$ of the performance variance is explained by the 
linear regression model, and the Pearson correlation coefficient is below $0.23$. On the contrary, the autocorrelation length is highly correlated with EA
performance.
The Pearson correlation coefficient is $0.82$, and $r^2=67.9 \%$ of variance is explained by the simple linear regression. The result of the real-world NROO 
problem with costly fitness function is in accordance with fundamental works in EA such as on the well-known NK-landscapes~: the problem difficulty and the 
performances are correlated to the ruggedness of the fitness landscapes. In contrast to the classical result obtained on the previous fundamental works
however, the more rugged the landscape, the better the performance of the parallel EA. Our first result shows that a fitness landscape approach could be used to
tune the parameters, but for highly selective parallel $(1+\lambda)$-EA with a large number of computing units, 
rugged landscapes should be preferred.

\section{Conclusions}
\label{sec:conclusions}

\vspace{-0.3cm}
In this paper, a real-world black-box combinatorial optimization problem with an expensive fitness function has been studied, and to solve it, an asynchronous
master-worker $(1+\lambda)$-EA running on a massively parallel architecture was used. The tough point of this exercise was the design of the
algorithm, and mainly the mutation parameters. To do so, a parametric study was launched, giving satisfactory results, but requiring a lot of resources.
In a second time, a fitness landscape analysis on this expensive problem showed that it is possible to tune the mutation parameters, and surprisingly, in the
case of a large scale computing environment, with a limited user computation time, the mutation parameters associated to the most rugged landscape are
relevant. It has then been possible to improve the considered criterion of almost $65\%$, meaning that on a given load-following transient, the operation of 
the core keep the axial power offset almost constant. This is encouraging for the following as some margins have been generated so that more heckled transients
can now be considered.

The next step of this work is the minimization of the rejected effluent by the nuclear power plant. While there are many steps to be taken, our methodology opens the opportunity to tune the evolutionary algorithm from fitness landscape features, and pushes to design an efficient bi-objective algorithm for combinatorial black-box problems with expensive fitness functions.

\bibliographystyle{abbrv}

\end{document}